\documentclass[final,3p,times,twocolumn]{elsarticle}
\usepackage{lineno,hyperref}
\modulolinenumbers[5]

\usepackage{siunitx}
\usepackage{textcomp} 
\DeclareSIUnit\permille{\text{\textperthousand}}
\journal{Nucl. Instrum. Methods Phys. Res. A}

\begin{document}

\begin{frontmatter}

\title{The DESY Digital Silicon Photomultiplier: Device Characteristics and First Test-Beam Results}

\author[]{Inge Diehl}
\author[]{Finn Feindt\corref{corresAuthor}}
	\cortext[corresAuthor]{Corresponding author}
	\ead{finn.feindt@desy.de}
\author[]{Karsten Hansen}
\author[]{Stephan Lachnit\fnref{uhh}}
\author[]{Frauke Poblotzki}
\author[]{Daniil Rastorguev\fnref{wupp}}
\author[]{Simon Spannagel}
\author[]{Tomas Vanat}
\author[]{Gianpiero Vignola\fnref{bonn}}
\address{Deutsches Elektronen-Synchrotron DESY, Notkestr. 85, 22607 Hamburg, Germany}

\fntext[uhh]{Also at University of Hamburg, Germany}
\fntext[wupp]{Also at University of Wuppertal, German}
\fntext[bonn]{Also at University of Bonn, Germany}

\begin{abstract}
Silicon Photomultipliers (SiPMs) are state-of-the-art photon detectors used in particle physics, medical imaging, and beyond. They are sensitive to individual photons in the optical wavelength regime and achieve time resolutions of a few tens of picoseconds, which makes them interesting candidates for timing detectors in tracking systems for particle physics experiments. The Geiger discharges triggered in the sensitive elements of a SiPM, Single-Photon Avalanche Diodes (SPADs), yield signal amplitudes independent of the energy deposited by a photon or ionizing particle. This intrinsically digital nature of the signal motivates its digitization already on SPAD level.
	
A digital SiPM (dSiPM) was designed at Deutsches Elektronen Synchrotron (DESY), combining a SPAD array with embedded CMOS circuitry for on-chip signal processing. A key feature of the DESY dSiPM is its capability to provide hit-position information on pixel level, and one hit time stamp per quadrant at a \SI{3}{\mega\hertz} readout-frame rate. The pixels comprise four SPADs and have a pitch of about \SI{70}{\micro\meter}. The four time stamps are provided by \SI{12}{bit} Time-to-Digital Converters (TDCs) with a resolution better than \SI{100}{\ps}.

The chip was characterized in the laboratory to determine dark count rate, breakdown voltage, and TDC characteristics. Test-beam measurements are analyzed to assess the DESY dSiPMs performance in the context of a 4D-tracking applications. The results demonstrate a spatial hit resolution on a pixel level, a minimum-ionizing particle detection efficiency of about \SI{30}{\percent} and a time resolution in the order of \SI{50}{\pico\second}.
\end{abstract}


\end{frontmatter}


	\section{Introduction}
Silicon Photomultipliers (SiPMs) are arrays of Single-Photon Avalanche Diodes (SPADs). These SPADS are planar pn-junctions enhanced by a thin layer with high doping concentration, resulting in a high electric field, such that Geiger-mode operation is feasible at reverse bias voltages of a few \SI{10}{\volt}. A Geiger discharge can be triggered by ionizing radiation or single photons in the optical wavelength regime, with detection efficiencies on the order of \SI{50}{\percent}. Due to the avalanche multiplication, the signal rise times of SiPMs are sufficiently short to reach a temporal resolution of a particles time-of-arrival on the order of few \SI{10}{\pico\second}. Comprehensive introductions into the topic are given in for example in~\cite{klanner,acerbi,wermes}.

Another feature of the avalanche multiplication is that the amplitude of the current induced by the avalanche is independent of the amount of charge carriers produced by a particle or photon interaction. This absence of charge information suggests the conversion into a binary signal, by means of a discrimination, without any loss of information. On the other hand, digital signal processing on a SPAD level adds a lot of benefits. For example it allows to extract the position of a firing SPAD within the SPAD array. Such digital SiPMs (dSiPMs), produced in a standard CMOS process, where first introduced in~\cite{frach}.

Applications for dSiPMs include the use of dSiPMs as light-sensitive sensor for the readout of scintillating fibers~\cite{fischer}. This work explores the possibility to use a dSiPM, designed at Deutsches Elektronen Synchrotron (DESY), as a 4D-tracking detector. In this case, a 4D-tracking device is to be understood as a detector providing the measurement of a Minimum-Ionizing-Particle's (MIPs) impact coordinates and time-of-arrival, as discussed in~\cite{nicolo}. The said dSiPM will be briefly described in the following section.

	\section{The DESY dSiPM}
\begin{figure}[tbp]
	\centering
	\includegraphics[width=0.33\textwidth]{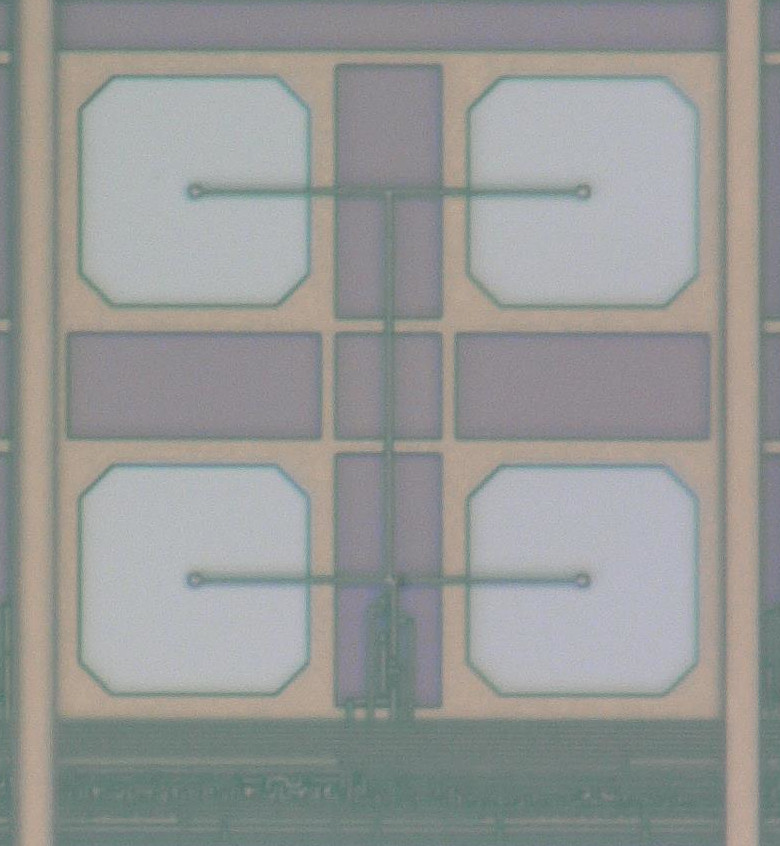}
	\caption[]{Microscopic image of a single pixel. The four squares with cut-off corners are the SPADs. The in-pixel circuitry is visible at the bottom of the image. Taken from~\cite{inge}.}  \label{pixel}
\end{figure}

\begin{figure*}[tbp]
	\centering
	\includegraphics[width=0.60\textwidth]{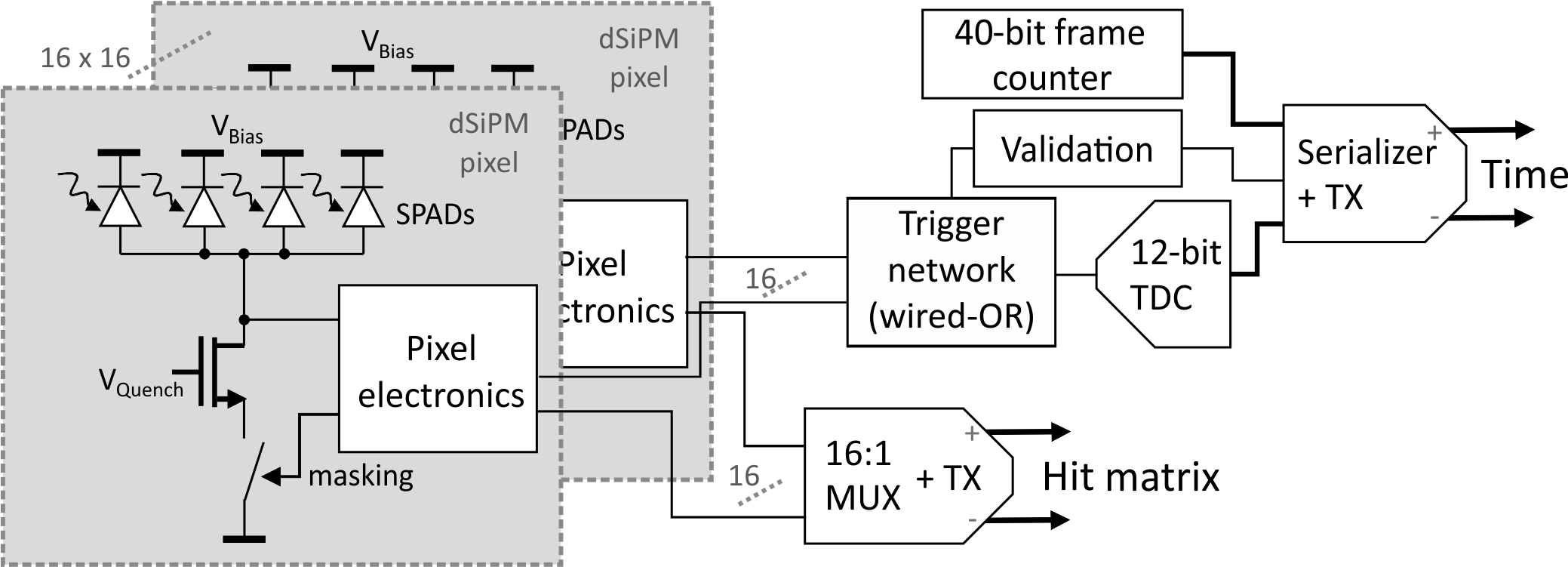}
	\caption[]{Simplified readout scheme of a quadrant of the DESY dSiPM.} \label{ro}
\end{figure*}

The sensor is designed and produced in the LFoundry 150-\SI{}{\nano\meter} CMOS technology. The key building blocks are SPADs, with a fixed layout provided by the manufacturer. Four of these SPADs, with an active area of $20 \times$\SI{20}{\square\micro\meter}, form a pixel with a size of $69.6 \times$\SI{76}{\square\micro\meter}, as shown in figure~\ref{pixel}. The fill factor of a pixel, and hence the entire matrix, amounts to \SI{30}{\percent}. This includes the readout circuitry implementing the key pixel functionality, i.e. a masking mechanism, a variable quenching resistance (transistor), a discriminator, a counter (up to \SI{2}{bit}), and a fast wired-OR connection to the quadrant's Time-to-Digital Converter (TDC). A quadrant is the next-larger building block of the chip. It comprises $16 \times 16$ pixels, and features a \SI{12}{bit} TDC with two stages, a validation logic, and multiplexers for the hit and timestamp information in the chip periphery. The TDC provides the time stamp of the first firing pixel within a \SI{333}{\nano\second} readout frame. An overview of the readout scheme is shown in figure~\ref{ro}, and a detailed description of the DESY dSiPM, including details on the circuit design and laboratory characterization, is given in~\cite{inge}.

For the reader's convenience, a few measurements of key properties of the DESY dSiPM from~\cite{inge} are summarized in the following. The sensors reach Geiger-mode operation at about \SI{18}{\volt}, and the temperature coefficient of this breakdown voltage is about \SI{16}{\milli\volt\per\kelvin}. The dark count rate strongly depends on temperature and over voltage, defined as the difference between the applied bias and the breakdown voltage. At a typical working point, e.g. an over voltage of \SI{2}{\volt} and a temperature of \SI{0}{\degreeCelsius}, it reaches \SI{7}{\kilo\hertz} per pixel, which corresponds to \SI{4.4}{\hertz\per\square\micro\meter} accounting for the active area. The time resolution of the TDC is reported to be $95.8 \pm$\SI{13.65}{\pico\second} (average bin width).

The readout system for the sensor is based on the Caribou system~\cite{vanat}. The main building block of the system is a System-on-Chip (SoC) board, with a CPU and an FPGA on the same die. The CPU runs the data-acquisition software, Peary~\cite{peary}, and the FPGA executes custom firmware blocks to handle configuration and readout of the chip. The Control and Readout (CaR) board provides all the peripherals required for the chip operation, and the chip itself is glued and wire bonded to a custom chip board with only few active components.
 	
	\section{Test-Beam Setup and Reconstruction}
\begin{figure}[tbp]
	\centering
	\includegraphics[width=0.499\textwidth]{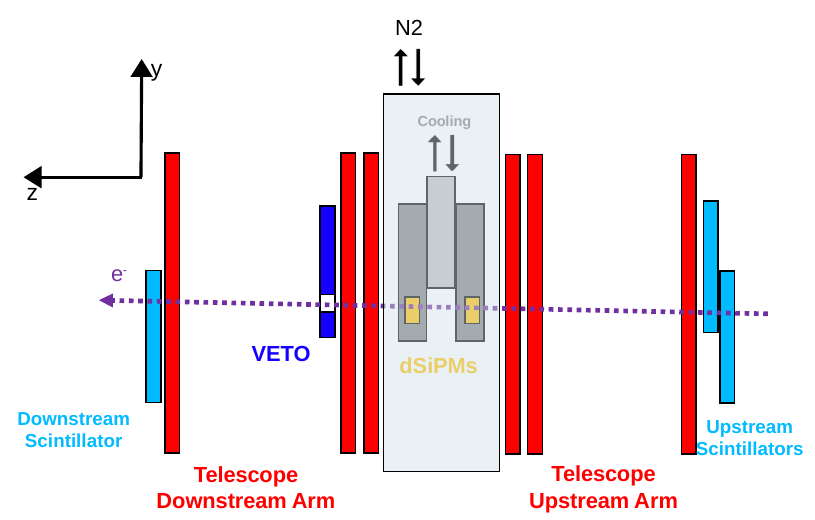}
	\caption[]{Schematic picture of the test-beam setup used for the presented measurements.} \label{setup}
\end{figure}

To explore the capabilities of the DESY dSiPM as a detector for ionizing radiation, test-beam measurements have been performed at the DESY II Test Team Facility~\cite{desyII}, which offers a beam of electrons or positrons with selectable momenta between \SI{1}{\giga\electronvolt\per c} and \SI{6}{\giga\electronvolt\per c}. A schematic picture of the setup is shown in figure~\ref{setup}. The main reference system is an EUDET-type beam telescope~\cite{eudet}. It consists of six tracking planes, each a MIMOSA-26 sensor~\cite{mimosa}, arranged in an upstream and downstream arm of three planes. Four plastic scintillators with photomultiplier readout are used to provide a trigger signal. Three of the scintillator signals are used in coincidence, while the 4th scintillator features a \SI{2}{\milli\meter} hole in its center and provides a veto signal, such that particles traversing the hole are most likely to cause a trigger signal. This setup maximizes the fraction of triggers where a particle traverses the active area of the DESY dSiPM, and yields a trigger rate on the order of \SI{100}{\hertz}. An  AIDA Trigger Logic Unit (TLU)~\cite{tlu} is used for the distribution of clock and trigger signals. Since it is not trivial to provide a time reference with sufficient precision, two DESY dSiPMs were installed, such that their time stamps can be compared. The chip boards with the DESY dSiPMs are placed inside an aluminum encasing, for physical and light protection, which is in contact with a cooling element allowing stable temperatures down to about \SI{-5}{\degreeCelsius}. Both DESY dSiPMs and their encasing share a PMMA box flushed with nitrogen to avoid condensation. The chip boards, the encasing, and the PMMA box feature windows to minimize multiple coulomb scattering. Thin aluminum and polyimide sheets are used to keep light-, and air-tightness, respectively.

The reconstruction framework Corryvreckan~\cite{curry,curry_web} is used for the analysis of the recorded data. The reconstruction chain starts with decoding of the beam-telescope and DESY dSiPM data. Events are defined based on the trigger IDs provided by the TLU. For each detector plane and event, hits in neighboring pixels are grouped to form clusters, and reference tracks are reconstructed from clusters in the beam-telescope planes using the General Broken Lines track model~\cite{gbl1,gbl2} as implemented in the framework. Hits in the DESY dSiPMs within a radius of about \SI{70}{\micro\meter} around a reconstructed track are associated to that track and used to study the MIP-detection efficiency, spatial, and temporal resolution.

	\section{Results}
\begin{figure}[tbp]
	\centering
	\includegraphics[width=0.499\textwidth]{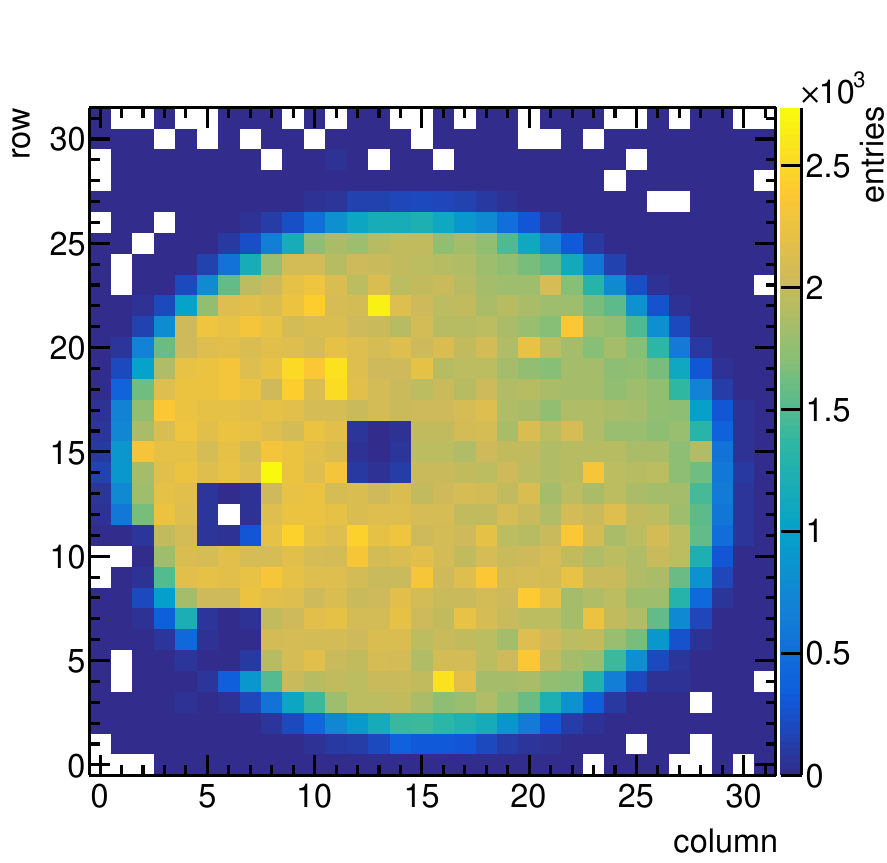}
	\caption[]{Intercept positions of associated tracks for a DESY dSiPM operated at a temperature of about~\SI{0}{\degreeCelsius} and an over voltage of \SI{2}{\volt}.} \label{hitMap}
\end{figure}

The reported results are based on data recorded at the DESY II Test Team Facility, using electrons with an energy of \SI{4}{\giga\electronvolt}, and provide an overview of the DESY dSiPM's key properties in terms of MIP-detection. The investigated DESY dSiPM is operated at a temperature of about~\SI{0}{\degreeCelsius} and an over voltage of \SI{2}{\volt}. Figure~\ref{hitMap} shows the intercept positions of associated tracks with the DESY dSiPM. The circular feature corresponds to the trigger acceptance, i.e. the hole of the veto scintillator, and the horizontal gradient within is due to the beam-intensity profile. The dark spots within the circle correspond to pixels that are masked due to an excessive dark count rate, to avoid biasing the efficiency measurements. The following results are constrained to the region within the trigger acceptance.

\begin{figure}[tbp]
	\centering
	\includegraphics[width=0.499\textwidth]{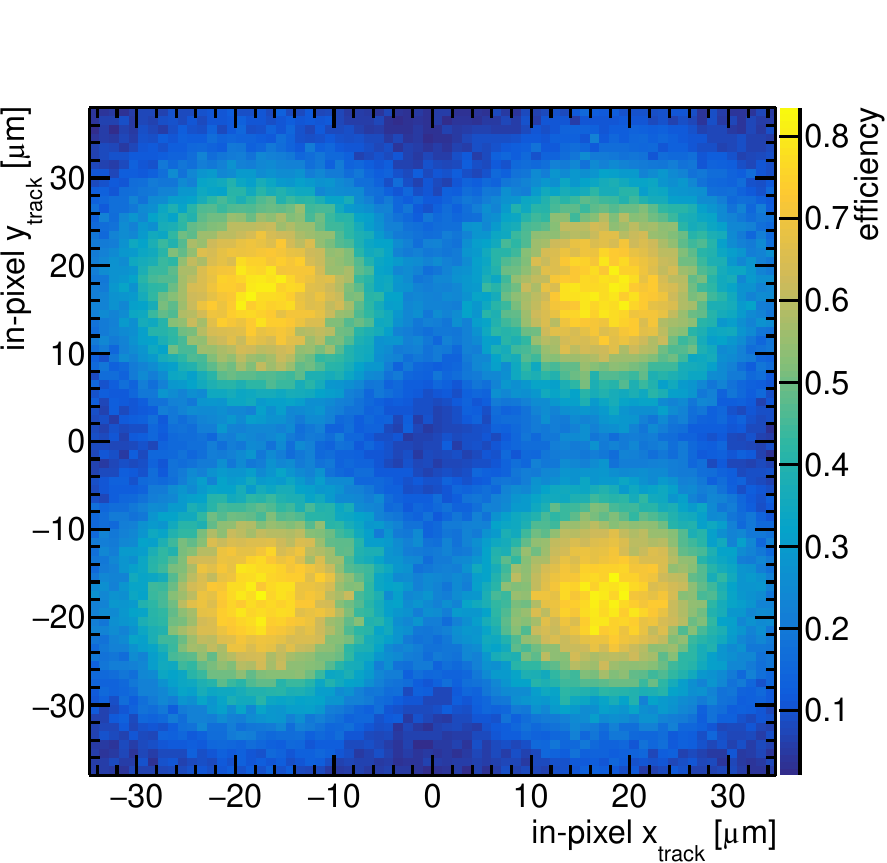}
	\caption[]{In-pixel efficiency map of a DESY dSiPM operated at a temperature of about~\SI{0}{\degreeCelsius} and an over voltage of \SI{2}{\volt}. Average over the unmasked pixels within the trigger acceptance. Compare to figure~\ref{pixel} for pixel layout.} \label{effInPix}
\end{figure}

An in-pixel efficiency map is obtained by projecting the reconstructed particle intercepts with the DESY dSiPM ($x_{track}$, $y_{track}$) into a single pixel and averaging the efficiency for the binned in-pixel positions. Such a map is shown in figure~\ref{effInPix}, the displayed area corresponds to the one shown in figure~\ref{pixel}. As can be seen, the DESY dSiPM is highly efficient when the electrons hit the SPADs, but inefficient in the area in-between. The geometric setup, due to an inhomogeneous distribution of material, does not allow estimating the track resolution at the position of the dSiPM. However, assuming a track resolution of about \SI{6.1}{\micro\meter}, the results are in agreement with the hypothesis of smeared-out position information and a fully efficient SPAD of the nominal area. This is supported by the fact that the measured total efficiency of about \SI{32}{\percent} agrees with the fill factor of the sensor. Systematic effects due to dead time and fake hits are estimated to be on the order of a few percent.
\begin{figure}[tbp]
	\centering
	\includegraphics[width=0.499\textwidth]{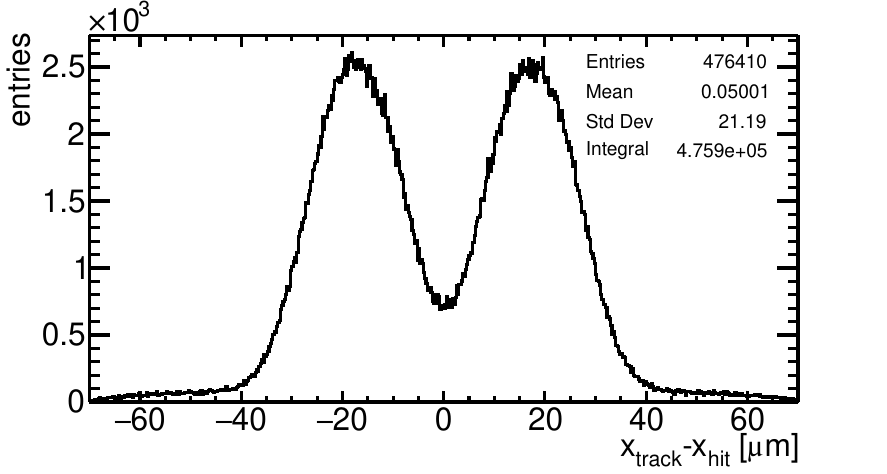}
	\caption[]{Residual distribution for a DESY dSiPM operated at a temperature of about~\SI{0}{\degreeCelsius} and an over voltage of \SI{2}{\volt}. The selection includes the unmasked pixels within the trigger acceptance.} \label{residuals}
\end{figure}

In figure~\ref{residuals}, the distribution of the residuals between $x_{track}$ and the position measurement provided by the DESY dSiPM itself ($x_{hit}$) is shown. For a fully efficient sensor with the same pitch, the core of the distribution would be expected to resemble a box, convolved with a Gaussian to account for the finite track resolution. Due to the inefficiency between SPADs, shown in figure ~\ref{effInPix}, a two-peaked structure appears instead. The background, visible for values below \SI{-40}{\micro\meter} and above \SI{40}{\micro\meter}, is due to an association of noise hits. The standard deviation of the distribution of about \SI{21}{\micro\meter} is within expectations for a sensor of the given pitch and confirms a spatial resolution on the level of a single pixel.
\begin{figure}[tbp]
	\centering
	\includegraphics[width=0.499\textwidth]{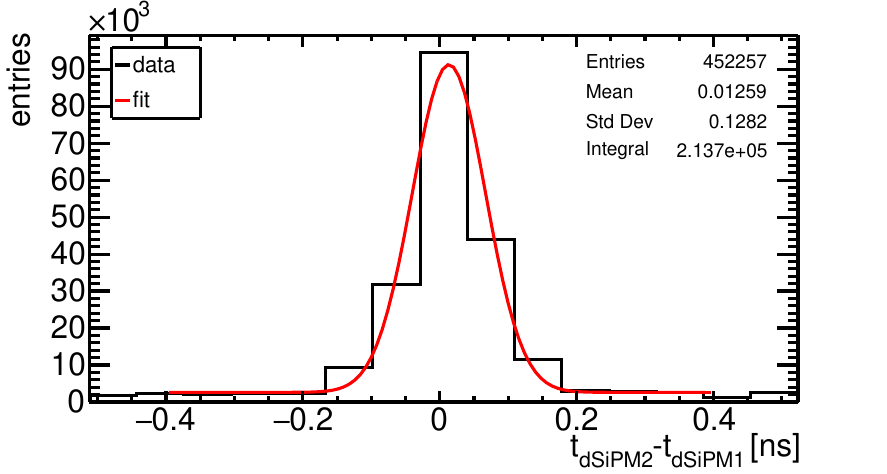}
	\caption[]{Time residual between the time stamps provided by the two installed DESY dSiPMs, operated at a temperature of about~\SI{0}{\degreeCelsius} and an over voltage of \SI{2}{\volt}.} \label{residuals_time}
\end{figure}

To study the time resolution of the DESY dSiPM, the residual between the time stamps provided by the two installed DESY dSiPMs is calculated. The second DESY dSiPM is operated at a similar temperature and over voltage, and referred to as dSiPM2. The time stamps are corrected for differential non-linearity in the TDCs. Corrections for propagation delays are expected to be small, due to the symmetry of the setup, and are not applied. The corresponding distribution is shown in figure~\ref{residuals_time}. It contains three contributions from both DESY dSiPMs, a fast and a slow response, background due to noise, and combinations of these, as detailed in~\cite{stephan}. The noise-background is flat, while the slow response causes tails on the scale of a nanosecond. The fast response contributes the central peak, which has a width of \SI{54}{\pico\second}, extracted fitting a Gaussian distribution and constant background, and contains about \SI{85}{\percent} of the data. This corresponds to a single DESY dSiPM contribution of about \SI{38}{\pico\second}. It should be noted that this result depends on the phase relation between the synchronized clocks of the two DESY dSiPMs. Systematic studies, including a proper description of all three components and different operation conditions, are in preparation and indicate that the fast component is generally below \SI{55}{\pico\second}~\cite{stephan}.

\begin{figure}[tbp]
	\centering
	\includegraphics[width=0.499\textwidth]{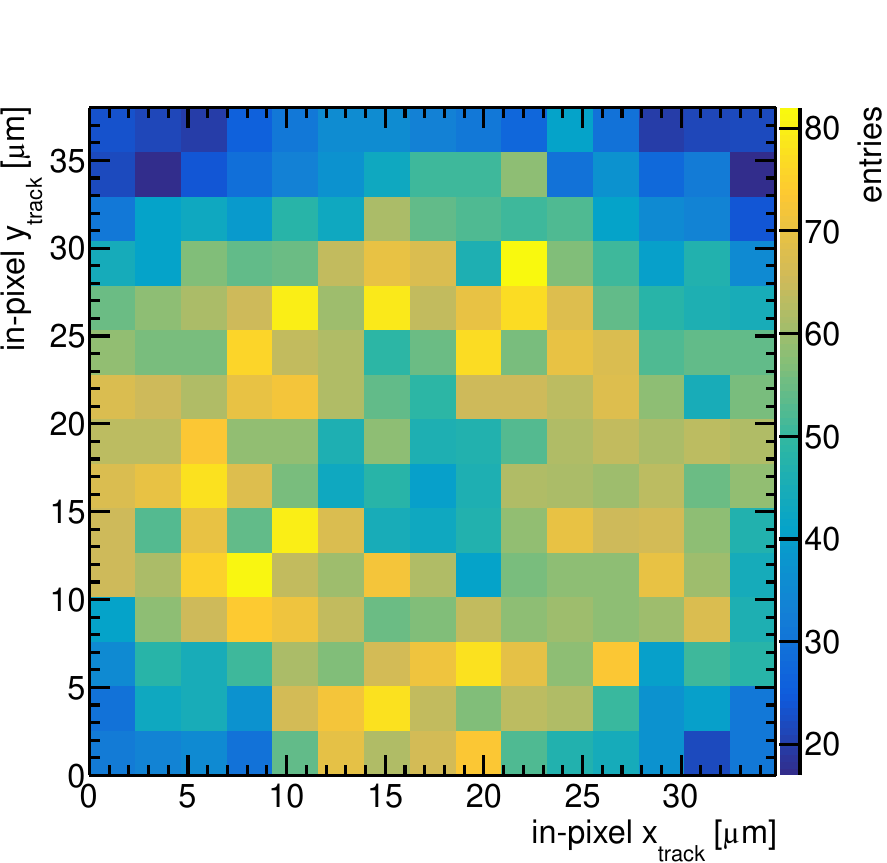}
	\caption[]{Intercept positions of tracks associated to events with a slow response. Folded into a single SPAD, due to limited statistics.} \label{rings}
\end{figure}

To investigate the origin of the slow response, these events are selected, and the intercept positions of the associated tracks are plotted in figure~\ref{rings} (note different axes with respect to figure~\ref{effInPix}). The ring-shaped feature indicates, that the slow signals are predominantly due to hits close to the edge of the SPADs. Potential causes for this are a lower electric-field strength in the edge-region, that comes with longer drift times into the multiplication region, a similar effect due to diffusion, or a combination thereof.

	\section{Conclusion and Outlook}
The shown results represent a first assessment of the MIP-detection capabilities of a digital silicon photomultiplier. The DESY dSiPM chip combines a CMOS SPAD design with integrated readout circuitry. It is successfully operated in synchronization with a system of reference detectors at the DESY II Test Team Facility, demonstrating the possibility of integration into complex detector systems. Temporal, and spatial resolution of the sensor are demonstrated to be on the order of \SI{50}{\pico\second}, and \SI{21}{\micro\meter}, respectively. The detection efficiency is measured to be on the order of \SI{30}{\percent}, in agreement with the fill factor. Further studies on the recorded data will investigate the dependence of these properties on operation conditions, such as over voltage and temperature, and the systematics of the data and reconstruction approach.

It is to be added, that the low measured hit-detection efficiency is probably insufficient for 4D-tracking applications. However, measurements in~\cite{franzi} show that the Cherenkov effect in the encapsulation of a SiPM can increase its MIP-detection efficiency. Studies on the effect of a thin radiator glued on top of a DESY dSiPM on the MIP-detection efficiency are planned. Also the temporal resolution is expected to improve in this scenario, as a larger number of photons increases the probability for a fast response from the SPAD center. On the other hand, these additional photons might have detrimental effects on the spatial resolution obtainable with such a dSiPM-radiator assembly. We are excited for the results of these studies.

	\section{Acknowledgments}
The authors would like to thank A. Venzmer, E. W{\"u}stenhagen and D. Gorski for test-board and mechanical case design, test setup, as well as chip assembly.

Measurements leading to these results have been performed at the Test Beam Facility at DESY Hamburg (Germany), a member of the Helmholtz Association (HGF).
\bibliography{mybibfile}

\end{document}